\documentclass[aps,prl,preprint,showpacs]{revtex4}

\usepackage{graphicx}
\usepackage{epstopdf}
\usepackage{dcolumn}
\usepackage{bm}
\usepackage{color}
\usepackage{natbib}
\usepackage{amsmath}
\usepackage{SIunits}

\begin{document}

\title{Quantification of Depletion Induced Adhesion of Red Blood Cells}

\author{P.~Steffen$^{1}$,C.~Verdier$^{2}$, C.~Wagner$^{1}$}
\affiliation{$^1$
Experimental Physics, Saarland University, 66041 Saarbreucken, Germany  \\
$^2$  CNRS - Universit\'e de Grenoble I, UMR5588, Laboratoire Interdisciplinaire de Physique (LIPhy)
38041 Grenoble, France}


\begin{abstract}
Red blood cells (RBC) are known to form aggregates in the form of rouleaux due to the presence of plasma proteins under physiological conditions. The formation of rouleaux can also be induced \textit{in vitro} by the addition of macromolecules to the RBC suspension. Current data on the adhesion strength between red blood cells in their natural discocyte shapes mostly originate from indirect measurements such as flow chamber experiments, but data is lacking at the single cell level. Here, we present measurements on the dextran-induced aggregation of red blood cells using atomic force microscopy-based single cell force spectroscopy (SCFS). The effects of dextran concentration and molecular weight on the interaction energy of adhering RBCs were determined. The results are in excellent agreement with a model based on the depletion effect and previous experimental studies.
\end{abstract}

\maketitle \vskip2pc

\begin{figure} [ht!]
\includegraphics[scale=0.3]{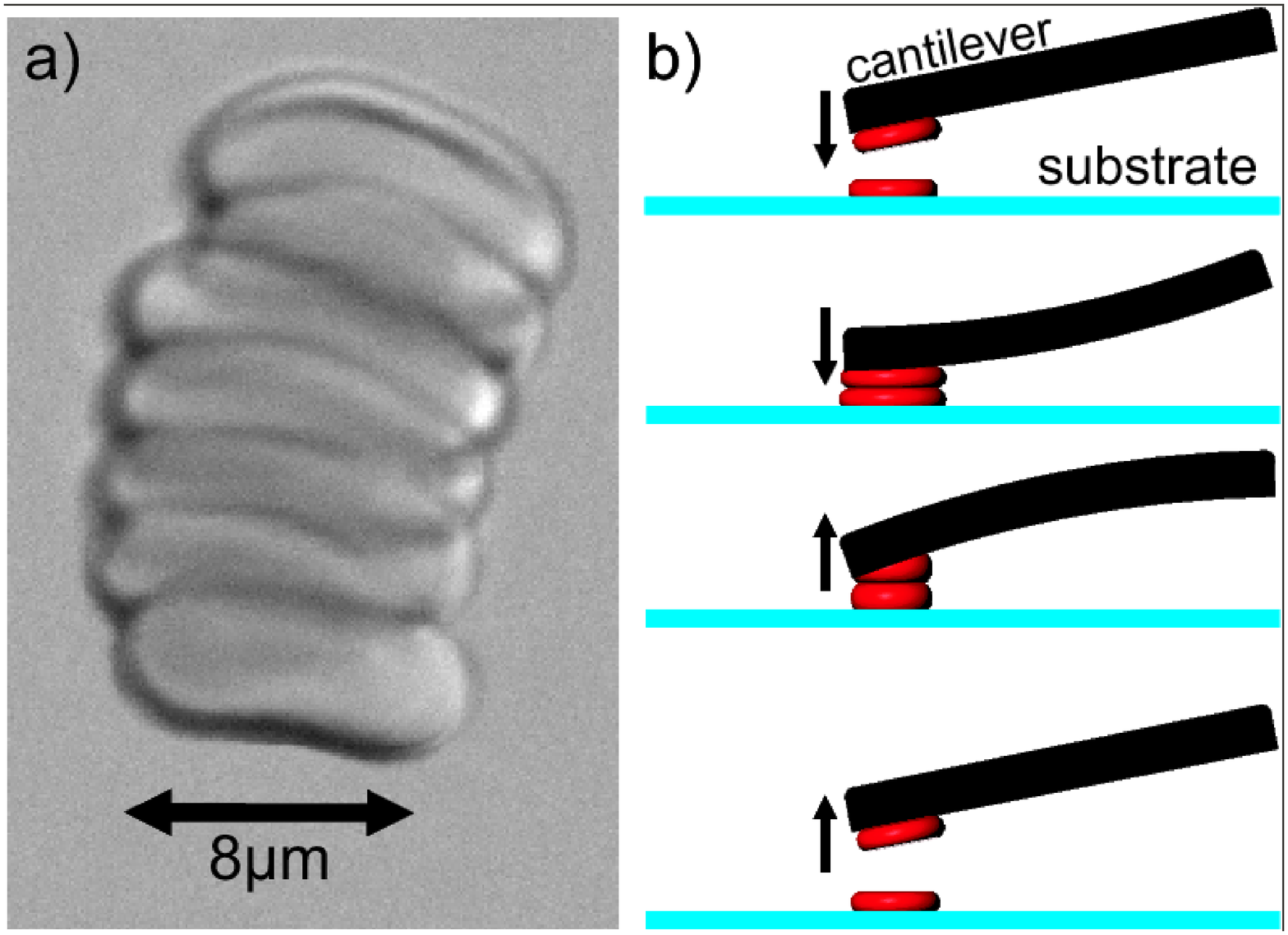}
\caption{a) Snapshot of a rouleau of 7 RBCs in a dextran solution. b) A sketch of the working principle of single cell force spectroscopy (SCFS). A single cell is attached to the pre-functionalized cantilever and is lowered onto another cell, which is fixed at the bottom. The adhesion force and adhesion energy are measured while withdrawing the cells.}
\label{Introduction}
\end{figure}

\begin{figure} [ht!]
\includegraphics[scale=0.8]{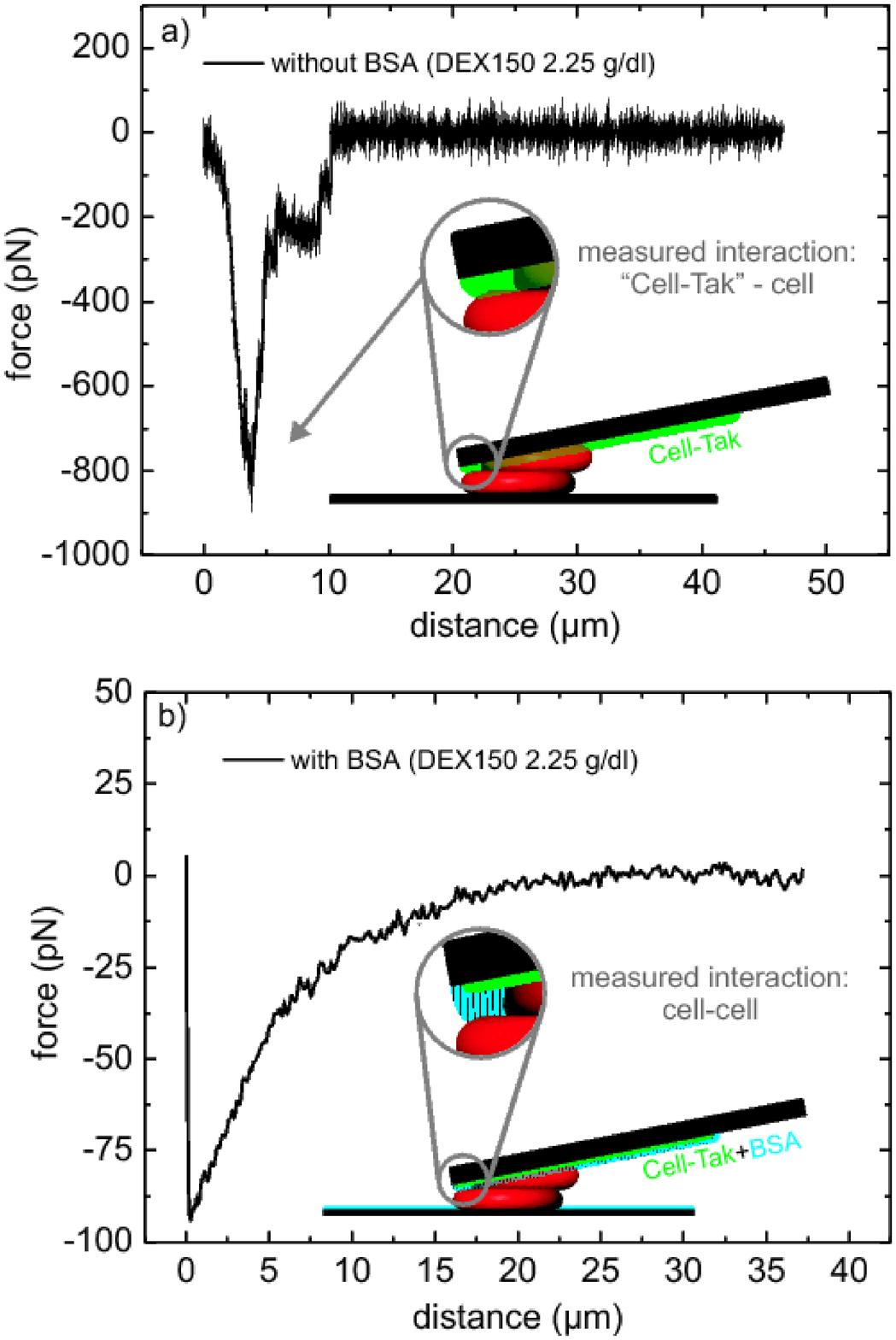}
\caption{a) and b) show the effects of BSA treatment (see the text for details). Without BSA treatment,  undesired adhesion events occur whose origin is not the investigated depletion effect; e.g., the cells don't hit concentrically and the lower cell touches the Cell-Tak (i.e., a stonger adhesion force is measured because of the strong adhesiveness of the Cell-Tak). With BSA treatment, the Cell-Tak is completely passivated and the influence of those undesired adhesion events is minimized, as the changes in shape and magnitude of the measured force curve document. }
\label{BSA}
\end{figure}

\begin{figure} [ht!]
\includegraphics[scale=0.7]{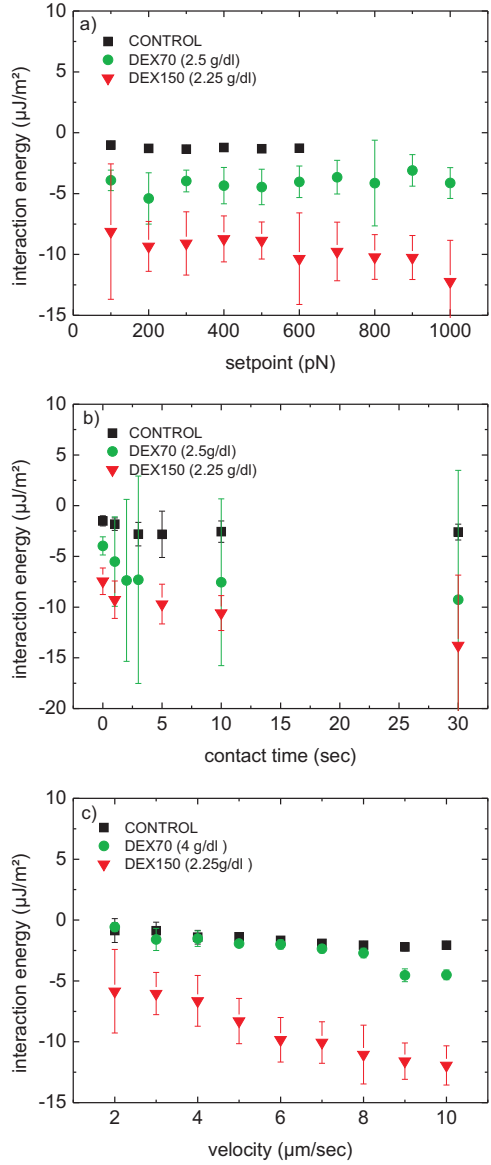}
\caption{Parameter measurements: a) shows the dependence of the measured adhesion energy on the chosen force setpoint $F_{set}$. In all measurements, no significant dependence on $F_{set}$ was observed. b) shows the dependence of the measured adhesion energy on the chosen contact time $\tau$ of both cells. Increasing contact time leads to an increase in interaction energy and error bars. c) shows the dependence of the measured adhesion energy on the chosen withdrawal velocity $v$ of the cantilever. At higher velocities in the DEX150 measurements a dependence on the cantilever velocity was observed, but for moderate velocities this dependence was still less than the error in the measurement. }
\label{parameters}
\end{figure}

\begin{figure} [ht!]
\includegraphics[scale=1.0]{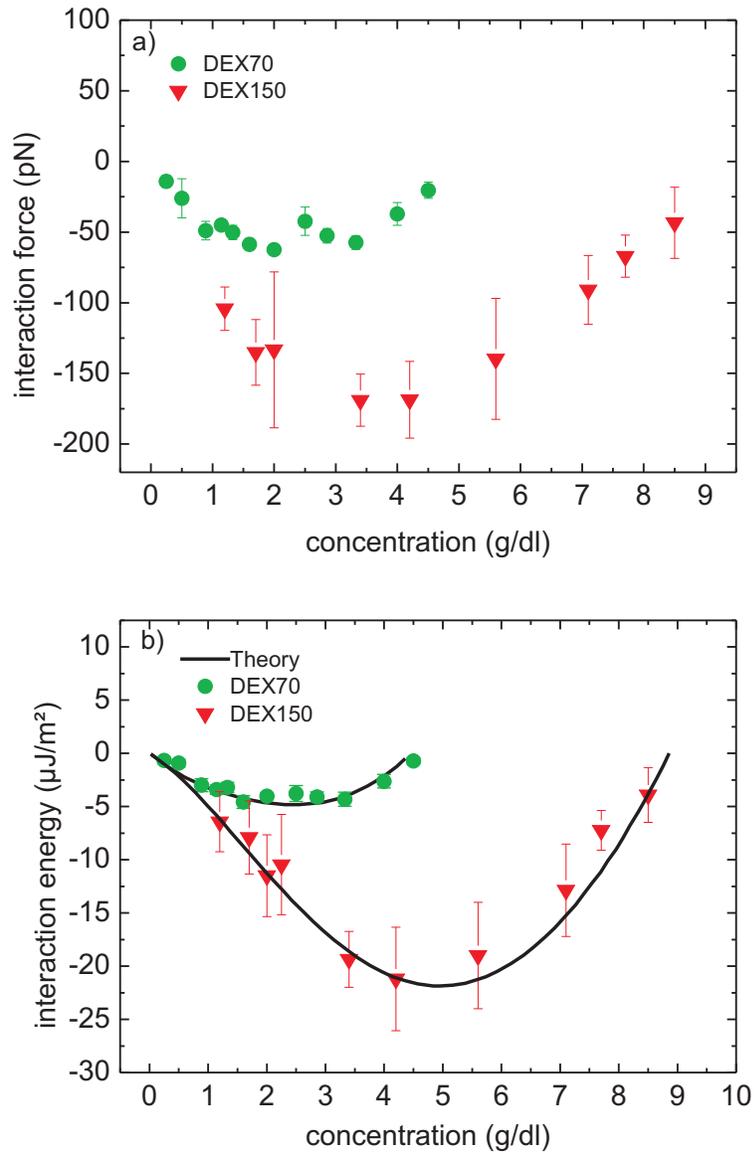}
\caption{a) The measured adhesion forces of the dextrans at different concentrations. The maximum interaction strengths were observed at $2 \gram/\deci\litre$ (DEX70) and $4 \gram/\deci\litre$ (DEX150). b) the dependence of the interaction energy of two red blood cells on the concentrations of two dextran types. The solid line represents the curve calculated by Neu et al. \citep{Neu2002}. }
\label{interaction_energy}
\end{figure}

Non-pathological aggregation of RBCs or "rouleaux formation" (Fig.\ref{Introduction}a)  \textit{in vivo} is frequently observed and caused by the fibrinogen in plasma \citep{Fahraeus1929}. These aggregates are reversibly formed and can be dispersed by moderate shear rates. Thus, the shear thinning viscosity of blood is determined to a large extent by the formation and breaking of these aggregates. Irreversible red blood cell aggregation could be a microcirculatory risk factor and indicative of disease because irreversible aggregates can be observed in diseases such as malaria, multiple myeloma, inflammation \cite{Rampling2004} or in pathological thrombus formation \cite{Steffen2011}. Reversible rouleaux formation can also be induced by resuspending the RBCs in physiological solutions containing neutral macromolecules such as dextran (DEX) \cite{Pribush2007}. The fibrinogen mediated aggregation of RBCs increases consistently with increasing fibrinogen concentration \cite{Marton2001}, whereas the dextran-mediated aggregation of RBCs reaches a maximum at a certain dextran concentration. The strength of the aggregation depends not only on the dextran concentration, but also on the molecular weight of the dextran (i.e., the radius of gyration of the dextran) \cite{Brooks1988,Chien1987}. Previously, two different models have been developped to explain the aggregation of RBCs in polymer solutions: the Cross-Bridging model and the Depletion model. The Bridging model has been favored for a long time and has been proposed for plasma protein and neutral dextran macromolecule-induced RBC aggregation \cite{Brooks1973}. In this model, it is assumed that fibrinogen or dextran molecules non-specifically adsorb onto the cell membrane and form a "bridge" to the adjacent cell. However, over the most recent decade, more and more evidence has been observed in favor of the depletion model \cite{Neu2002,Neu2006,Meiselmann2009,Baeumler1996,Armstrong1999}. A first explanation of depletion forces was given by Asakura and Oosawa \cite{Asakura1958}, who discovered that the presence of small spheres (i.e., macromolecules) can induce effective forces between two larger particles if the distance between them is small enough. The origin of these forces is purely entropic. If the distance between the two large particles decreases to less than the size of the surrounding macromolecules, these macromolecules are expelled from the region between the particles. Consistently, the concentration of macromolecules becomes depleted in this region compared to that of the bulk, and an effective osmotic pressure causing an attraction between the large particles occurs. Neu et al. \cite{Neu2002} applied this concept of depletion-induced adhesion to red blood cells and developped a theoretical description of the interaction energy between two red blood cells in terms of the molecular weight and concentration of the dextran used. These results were compared with former measurements reported by Buxbaum et al. \cite{Buxbaum1982} based on the micropipette aspiration technique. Neu et al. were able to adapt their model parameters consistently to the experimental data of Buxbaum et al., but the number of data points remained limited.
In the present study, we used the technique of single cell force spectroscopy (SCFS) (Fig.\ref{Introduction}b) to measure the interaction energies between human red blood cells as functions of the molecular weight and concentration of dextran at the single cellular level and compared them to the predicted values of Neu et al.\cite{Neu2002}. Similar to them we used dextran70 (DEX70 with a molecular weight of $70 \kilo\dalton$) and dextran150 (DEX150 with a molecular weight of $150 \kilo\dalton$) from Sigma-Aldrich. In SCFS a single RBC is attached to a cantilever, while another cell is attached to the surface of the petri dish. Both cells are now brought into close contact and during the withdrawing of the cantilever the adhesion force between both cells is measured via the deflection of the cantilever (see Fig.\ref{Introduction}b). An atomic force microscope (AFM) (Nanowizard 2, equipped with the CellHesion Module with an increased pulling range of $100 \micro\meter$, JPK Instruments, Germany) was used to conduct single cell force spectroscopy measurements \cite{Friedrichs2010}. The spring constant of the cantilever was determined using the common thermal noise method (the cantilevers used were MLCT-O cantilevers with spring constants on the order of $0.01 \newton/\meter$, Bruker). Fresh blood from healthy donors was taken using a finger prick. The blood was obtained within one day of the experiment. The cells were washed three times by centrifugation ($2000 \gram, 3 \minute$) in a phosphate-buffered solution of physiological ionic strength. In the course of the experiment, a single RBC was attached to an AFM cantilever by appropriate functionalization. Cell Tak$^{TM}$ (BD Science) was used to bind a cell to the cantilever. A protocol was used, in which the cantilever was incubated in a Cell Tak$^{TM}$ drop. After $2\minute$, the Cell Tak$^{TM}$ solution was carefully removed; this was followed by a $3 \minute$  waiting period to allow the acetic acid from the Cell-Tak$^{TM}$ to evaporate from the cantilever. Rinsing the cantilever with ethanol and PBS (Phosphate-Buffered Saline, $137\milli M/\litre\;\; NaCl$, $2.7\milli M/\litre\;\; KCl$, $10\milli M/\litre\;\; Na_2HPO_42H_2O$, $2\milli M/\litre\;\; KH_2PO_4$, $pH=7.4$) completed the functionalization protocol. To attach a RBC to the cantilever, the latter was lowered manually until a preset cantilever deflection (i.e. force setpoint $F_{set}$) was reached, indicating contact between the cantilever and the cell. The cantilever was withdrawn continuously at low speed until the cell was no longer in contact with the surface. After the cell had been attached to the cantilever, the cell was placed on top of another cell that lay on the bottom of the petri dish. Functionalization of the plastic petri dish (PS, Polystyrol), to immobilize the bottom cell, was not necessary because RBCs adhered to the surface without any further treatment. While withdrawing (or retracting) the cantilever the adhesion force and adhesion energy were measured. The retraction curve is typically characterized by the maximum force required to separate the cells from each other and adhesion energy is calculated by computing the area under the retraction curve of the force distance curve. The interaction energies (or, more precisely, the interaction energy densities of two RBCs) are calculated by dividing the measured adhesion energies by the contact areas of the adhering cells using a value of $50.24 \micro\meter^2$ derived from the mean radius of RBCs.
The present study is exclusively concerned with adhesion that is caused by the presence of dextran molecules in the solution. Hence, any further source of adhesion (e.g., adhesion of the lower RBC to the Cell-Tak coated surface of the cantilever, see Fig.\ref{BSA}a) had to be excluded. For larger and stiffer cells, compared to RBCs, such undesired adhesion events are rarely observed, but for RBCs that have a height of just $2 \micro\meter$ this is an experimental difficulty; even with optimum (concentric) alignment, such binding to the surface was often observed in our first experiments. An example of those undesired adhesion events is shown in Fig.\ref{BSA}a. The measured adhesion forces were much higher than any reasonable estimate for dextran-induced adhesion. To overcome this problem, $0.1 \gram/\deci\litre$ BSA (Bovine Serum Albumin) was added to the solution after attaching the RBC to the cantilever. BSA could potentially induce an additional depletion interaction; however because the radius of gyration of BSA is only $3 \nano\meter$ \cite{Santos2003} and the concentration is fairly small, the additional depletion interaction due to the presence of BSA can be neglected. The effects of BSA treatment on RBCs (e.g., on cell geometry or mechanical properties) have been studied intensively \cite{Khairy2008,Jay1975,Williams1973}. However, any effect of the BSA treatment on the measured interaction energy can be neglected for the investigated adhesion because this adhesion is purely physical (i.e., we assume that there no adhesion proteins are involved that could possibly be blocked by the incorporation of BSA into the RBC membrane). In agreement with the literature, we found that in most of the cases, both the cells on the petri dish as well as on the cantilever remained in their physiological, discocyte shape.
The only purpose of the BSA is to passivate the surfaces of the cantilever and the petri dish. Thus, only the RBC surfaces contribute to the measured adhesion force arising from the depletion effect. Fig.\ref{BSA}b shows an example force curve after BSA treatment of the cantilever and the petri dish. The shape and magnitude of the force distance curve changes significantly. The extraordinary flexibility of RBCs allows them to stay in contact over large withdrawing distances after adhering, as one can see in the measurements with BSA (Fig.\ref{BSA}b). In the next step, the parameter setpoint force $F_{set}$, cantilever velocity $v$, and contact time $\tau$ of the cantilever were adjusted. Fig.\ref{parameters}a shows the interaction energy as a function of $F_{set}$ for DEX70, DEX150 and a measurement without dextran (control). The influence of the choice of $F_{set}$ on the measured interaction energy is negligible and was set to $F_{set}=300\pico\newton$ for the remaining measurements.
We will show below that we can describe our data well with a depletion model, but it is also known that macromolecular "bridging" between the RBCs can occur when the cells are in contact for a longer time \cite{Bronkhorst1997}. Bronkhorst et al.\cite{Bronkhorst1997} discovered that the time constant for those possible cross bridges is on the order of seconds. We varied the contact time from $\tau=0\second$ to $\tau=30\second$ to resolve the influence of the parameter $\tau$ (Fig.\ref{parameters}b). Large contact times lead to increased adhesion energies and increased error bars. Both can be indications for bridging events. Therefore, we attempted to minimize the contact time by setting $\tau$ to $0\second$. Depending on the cantilever velocity, the RBCs will be in "physical" contact for a longer time because due to their flexibility, the RBCs can stay in contact over distances of several $\mu\meter$. Hence, to minimize the actual contact time (to exclude bridging effects), the cantilever velocity had to be sufficiently high. This ensures that the measured interaction energies are purely depletion-induced.
Fig.\ref{parameters}c shows the dependence of the cantilever velocity $v$ on the measured interaction energies. In the control measurements, no influence of the velocity could be seen. On the contrary, a dependence of the velocity could be observed in the dextran measurements. We do not have a conclusive explanation for this dependence, but we assume that higher interaction energies lead to larger viscoelastic effects while deforming the RBCs. Due to the higher velocity, the RBCs are deformed to greater extent, and this might lead to higher apparent interaction energies. Up to a certain velocity, the effect of velocity can be neglected; e.g., for DEX70 this effect begins at velocities only higher than $9 \mu\meter/\second$. At high velocities, in the DEX150 measurements, this effect can be significant, while for moderate velocities, this effect is still less than the error measurement. As mentioned above, it is necessary to minimize the contact time such that any bridging effects can be excluded; i.e., the cantilever velocity must not be too small. Therefore, in all measurements, the cantilever velocity was chosen as $v=5 \micro\meter/\second$.\\
Fig.\ref{interaction_energy} shows the dependence of the adhesion force and the interaction energy on the dextran concentration. Each data point represents an average of 100 force curves for the same cell. The measured interaction energies are in excellent agreement with the predicted interaction energies given by Neu et al. \cite{Neu2002},who used an analytical approach to calculate the depletion interaction energy $E_D$ between two RBCs \cite{Neu2002}:
\begin{equation}
E_D=-2\pi(\Delta-d/2+\delta-p)
\end{equation}
where $\pi$ is the osmotic pressure, $\Delta$ is thickness of the depletion layer, d is the separation distance between adjacent surfaces, $\delta$ is the glycocalyx thickness and p is the depth of polymer penetration into the glycocalyx.
Their model combines electrostatic repulsion due to RBC surface charge and osmotic attractive forces due to polymer depletion near the RBC surface. The theory considers the soft surfaces of RBCs and the subsequent penetration depth $p$ of polymers into the surface. This penetration depth $p$ depends on the polymer type, concentration, and molecular size and is expected to be larger for small molecules and to increase with increasing polymer concentration due to increasing osmotic pressure $\pi$. With increasing osmotic pressure, the penetration of macromolecules into the soft RBCs deepens, impeding the depletion of macromolecules between both cells and hence reducing the interaction energy. Above a threshold concentration this effect becomes dominant, and this decreases the interaction energy, even though the concentration of macromolecules is increasing further. Taking the depletion effect and the soft surfaces of the RBCs into account, a bell-shaped dependence of the interaction energy on the dextran concentration was calculated (the solid line in Fig.\ref{interaction_energy}b).\\
In conclusion, we have presented single cell force spectroscopy measurements on dextran-induced red blood cell aggregation. The presence of dextran mimics the plasma molecules that lead to the formation of rouleaux under physiological conditions. Our findings are in excellent agreement with previous studies \cite{Neu2002}, and they can be described by a model based on the depletion effect. For contact times longer than a few seconds we find a slight tendency towards stronger adhesion energies. We can not conclusively decide whether this is due to bridging of the macromolecules between two cells or due to some other effect. One would need a sideview \cite{Canetta2005} of the adhesion areas while performing the adhesion tests to analyze this effect in greater detail.

\section{Acknowledgements}
This work was supported by the German graduate school GRK 1276 and the Nanoscience Foundation. We thank JPK Instruments for preliminary help with the experiments.

\section{Figure captions}

\end{document}